\documentclass[preprint,aps,12pt,showpacs,nofootinbib,tightenlines]{revtex4}

\usepackage[section]{placeins}
\usepackage{graphicx}
\usepackage{dcolumn}
\usepackage{bm}

\usepackage{epsfig}
\usepackage{amssymb}
\usepackage{amsmath}
\usepackage{flafter}
\usepackage{array}

\usepackage[dvips,usenames]{color}

\newlength{\dinwidth}
\newlength{\dinmargin}
\begin{document}

\title{A Three-Flavor AdS/QCD Model with a
Back-Reacted Geometry}
\author{ Yue-Liang Wu}
\email{ylwu@itp.ac.cn}
\author{Zhi-Feng Xie}
\email{xzhf@itp.ac.cn}
\affiliation{ Kavli Institute for Theoretical Physics China, Institute of Theoretical Physics, \\
Chinese Academy of Science (KITPC/ITP-CAS), Beijing, 100080,
P.R.China}

\begin{abstract}
A fully back-reaction geometry model of AdS/QCD including the
strange quark is described. We find that with the inclusion of the
strange quark the impact on the metric is very small and the final
predictions are changed only negligibly.
\end{abstract}

\maketitle

\thispagestyle{empty}

\newpage
\pagenumbering{arabic}

\section{Introduction}
QCD~\cite{first paper in qcd} is considered to be a well-established
theory for the strong interaction. In the high energy regime, we can
use a perturbative approach to understand the theory. However, at
low energy, because of the large coupling constant, perturbation
theory is not applicable. In the low energy regime we can appeal to
other methods of analysis, for instance chiral perturbation theory
and lattice QCD.

Conjectured by Maldacena \cite{Maldacena} in 1997, the AdS/CFT
correspondence is a new approach to this difficult problem. This
conjecture states that a string theory on $AdS_{5}\times S^{5}$ is
equivalent to a conformal theory on the boundary of $AdS_{5}$. QCD
is classically but not quantum mechanically conformal. However, the
AdS/CFT correspondence has provided important insights into QCD,
such as confinement at large distances \cite{Polchinski} and chiral
symmetry breaking\cite{Babington}, \cite{Mateos}, \cite{Erlich},
\cite{Hong:2004sa}, \cite{DaRold:2005zs}, \cite{Hirn:2005vk},
\cite{Ghoroku:2005vt}, \cite{wufengshock}. Currently these topics
are very active areas of research.

The quantitative correspondence was specified in independent work by
Gubser, Klebanov and Polyakov~\cite{Gubser} and by
Witten~\cite{Witten}
\begin{equation}
\left<e^{i\int d^4x\,{\cal O}\,\phi}\right>_{CFT}={\cal
Z}_{SUGRA}\left(\phi(z)|_{z\rightarrow 0} =\phi\right)
\end{equation}
which states that the generating functional for correlation
functions with a source $\phi$ for some field theory operator is
equivalent to the partition function of a supergravity theory where
the boundary value of some supergravity field is the source for the
field theory operator. The choice of supergravity field and field
theory operator is a matter of matching the representations of the
global symmetries of the two pair.

In recent years a new phenomenological approach, based on the rules
of the AdS/CFT correspondence has been developed
\cite{Erlich},\cite{DaRold:2005zs}. This approach introduces a
five-dimensional classical theory in an $AdS_5$ background where
appropriate fields are included in the action to act as sources on
the boundary for operators of a QCD-like theory. This original
formulation included only light quark operators and gave a
phenomological model of chiral symmetry breaking. In the
five-dimensional theory the symmetry is a gauge symmetry and a
simple Higgs mechanism is set up to model chiral symmetry breaking
in the four-dimensional theory.

The results from these relatively simple and phenomenological models
are remarkable and the simplest realisation gives postdictions for
several meson masses and decay constants with an average of around
$15\%$ error.

Since the introduction of this phenomenological action, many
advances have been made to model QCD more accurately. These include
the introduction of linear confinement via an appropriately chosen
scalar field in the five-dimensional theory\cite{Karch:2006pv}, the
inclusion of gluon condensate contributions to QCD
quantities\cite{Csaki:2006ji} and studies of heavy quark
potentials\cite{Boschi-Filho:2005mw},\cite{Boschi-Filho:2006pe},\cite{White:2007tu}.

In \cite{back reacted}, we considered the impact of a classical
scalar field back-reacting on the geometry. In this case the impact
on the geometry was most strongly affected by the condensate of
light quarks.

The strange quark was introduced \cite{wufengshock} in order to
study the kaon sector and found that reasonably accurate predictions
could be found for these mesons, too.

In this paper we ask what the impact of the strange quark on the
geometry will be. We may expect that as the chiral symmetry is
broken more explicitly for the strange quark the effect of the
strange quark condensate on the dynamics of the theory may be less
pronounced.

\section{Back Reaction on the Geometry}\label{sec.br}

In this section, we consider the impact of one scalar field on the
metric. The total field content is the gravitational field plus the
scalar field, which will be responsible for chiral symmetry
breaking. The Lagrangian is given by
\begin{equation}
\label{action} S= \int d^5x\sqrt{g}(-R+Tr(\partial\phi)^2+V(\phi)),
\end{equation}
$R$ is the five-dimensional Ricci scalar, and the metric is
\begin{equation}
ds^2= e^{-2A(y)} d x_{\mu} d x^{\mu}-dy^2.
\end{equation}
The Ricci scalar $R$ is given by
 \begin{equation}\label{ricci}
 R(y)=20A^{'2}(y)-8A^{''}(y)\ .
 \end{equation}
From the action, one can find the equations of motion for the scalar
field and for the metric tensor.
\begin{equation}
{1\over 2} g_{PQ} [-R+ Tr(\partial_{M} \phi \partial^{M} \phi +
V(\phi))] +R_{PQ} -Tr\partial_{P} \phi \partial_{Q} \phi =0\ ,
\end{equation}
and
\begin{equation}
Tr{\partial V(\phi) \over \partial \phi} = {2 \over \sqrt{g}}
Tr\partial_{P} ( \sqrt{g} g^{PQ} \partial_{Q} \phi )\ ,
\end{equation}
which gives
\begin{equation}
\label{th5}6A^{''}(y) -12 A^{'2}(y) + Tr(V(\phi) -\phi^{'2}(y)) =0\,
\end{equation}
\begin{equation}
 \label{th6}12 A^{'2}(y) - V(\phi) =Tr \phi^{'2}(y)\,
\end{equation}
\begin{equation}
 \phi^{''}(y) + 4A^{'}(y)\phi^{'}(y)+{1\over {2}}{\partial V(\phi) \over \partial \phi} =
 0.
\end{equation}
Eq.(\ref{th5}) and Eq.(\ref{th6}) give
\begin{equation}
\label{th8} 3A^{''}(y)=Tr\phi^{'2}(y)\,
\end{equation}
and
\begin{equation}
\label{th9} 3A^{''}(y)-12 A^{'2}(y) + V(\phi)=0\ .
\end{equation}
 ~From Eq.(\ref{th8}), the function of $A(y)$ can be obtained, given a solution for $\phi$. Then from
 Eq.(\ref{th9}), one can find the potential $V(\phi)$. So, at no
 point do we need to rely on numerical techniques.

We now give an example and show how to find the warp factor in the
metric function in the presence of a scalar field. Consider a scalar
field, given
by
\begin{equation}\label{eq.scalar}
\phi(y)=\frac{m_{q}}{2}e^{y}+\frac{\sigma}{2}e^{3y} \ ,
\end{equation}
where $~m_{q}$ and~$\sigma$ are 3 by 3 matrix.
\begin{equation}
m_{q}=diag(m,m,m_s),~~\nonumber\\
\sigma=diag(c,c,c_s).
\end{equation}
We find
\begin{equation}
 Tr\phi^{'2}(y)=2\big(\frac{3}{2}c e^{3y}+\frac{1}{2}m
 e^{y}\big)^{2}+\big(\frac{3}{2}c_s e^{3y}+\frac{1}{2}m_s
 e^{y}\big)^{2}.
\end{equation}
 Then from Eq.(\ref{th8}), and the UV boundary condition $A^{'}(y)_{y\to{-\infty}}=1$, the warp factor
 $A(y)$ is found to be
\begin{equation}\label{Ay}
A(y)=y+\frac{1}{8}\left(\frac{1}{3}c^{2}e^{6y}+\frac{1}{6}c_s^{2}e^{6y}+\frac{1}{2}c
e^{4y}+\frac{1}{4}c_s e^{3y}m_s\right).
\end{equation}
We see that the UV behaviour of the metric is not greatly modified
by the back reaction of this scalar field.
\section{A Phenomenological Model}

 In this model of QCD, the three relevant
operators are $~\bar{q}_{L}^{\alpha}q_{R}^{\beta}$,$~\bar{q}_{L,R}\gamma^{\mu}t^{a}q_{L,R}$.\\
 Now, let's consider the following action
 \begin{equation}\label{phenomenaction}
 S=\int d^5 x \sqrt{g} \left\{ -R + Tr \left( |D\phi|^2
+V(\phi)-\frac{1}{4g_{5}^{2}}(F_{L}^{2}+F_{R}^{2}) \right) \right\}
\end{equation}
where $D_{\mu}\phi=\partial_{\mu}\phi-iA_{L\mu}\phi+i\phi
A_{R\mu},A_{L,R}=A_{L,R}^{a}t^{a}$ and $
F_{\mu\nu}=\partial_{\mu}A_{\nu}-\partial_{\nu}A_{\mu}-i[A_{\mu},A_{\nu}]
$. We define the vector and axial-vector gauge bosons to be
$V_M={1\over2}(L_M+R_M)$ and $A_M={1\over2} (L_M-R_M)$ respectively.
Following \cite{Erlich} we choose the $V_{z}=A_{z}=0~$gauge. From
this action the classical value of the scalar field is found to be
that we chose in Eq.\ref{eq.scalar}. Substituting
$\phi=<\phi>e^{i2t^{a}\pi^{a}(x,y)}$ back into the action, the mass
matrix of $V_{M}$ and $A_{M}$ bosons can be calculated
\cite{wufengshock}.
\begin{equation}
M_{V}^{2}=\left(
\begin{array}{lll}
 \mathbf{0_{3 \times 3}} &0 &0 \\
 0 & {1\over 4} \left(\hat{m}-\hat{m}_s\right)^2 z^2 \mathbf{1_{4\times 4}} & 0 \\
 0& 0 & 0
\end{array}
\right),
\end{equation}
and
 \begin{equation}
M_{A}^{2}=\left(
\begin{array}{lll}
 \hat{m}^2 z^2 \mathbf{1_{3 \times 3}} &0 &0 \\
 0 & {1\over 4} \left( \hat{m}+\hat{m}_s \right)^2 z^2 \mathbf{1_{4\times 4}} & 0 \\
 0& 0 & {1\over 3} \left(  \left( \hat{m}\right)^2+2\left( \hat{m}_s \right)^2 \right)z^2
\end{array}
\right),
\end{equation}
where $\hat{m}=m+c z^2$ and $\hat{m}_s=m_s+c_s z^2$. The equations
of motion for the vector and axial vector bosons can also be derived
from Eq.(\ref{phenomenaction}). For convenience we make the
following change of variable $z=e^{y}$.
\begin{equation}
\left[ \partial_{z}^2 +\partial_{z}\left(\ln
a\right)\partial_z+\left( q^2 -( g_{5}^{2} a^2 M_{V}^{2})_{\alpha
\alpha} \right) \right] \Phi_{V}^{\alpha}(q,z)=0,\label{eq.phiV}
\end{equation}
and
\begin{equation}
\left[ \partial_{z}^2 +\partial_{z}\left(\ln
a\right)\partial_z+\left( q^2 -( g_{5}^{2} a^2 M_{A}^{2})_{\alpha
\alpha} \right) \right] \Phi_{A}^{\alpha}(q,z)=0,\label{eq.phiA}
\end{equation}
 where
 $a=a(z,m,c,m_s,c_s)=\frac{1}{z}exp\left(-\frac{1}{8}(\frac{1}{3}c^{2}z^{6}+\frac{1}{6}c_s^{2}z^{6}+\frac{1}{2}c
 z^{4}m+\frac{1}{4}c_s z^{4} m_s)\right)$,
with boundary conditions
$~\partial_{z}\Phi_{V}^{\alpha}(q,z_{IR})=0$, and
$~\Phi_{V}^{\alpha}(q,\epsilon)=0$, similarly
for$~\Phi_{A}^{\alpha}$. The mass of the vector and axial vector
mesons can be obtained by solving the eigenvalue equations
Eq.(\ref{eq.phiV}) and Eq.(\ref{eq.phiA}) with $q^{2}=m_{V}^{2}$ and
$q^{2}=m_{A}^{2}$, respectively. The decay constants of these mesons
can be obtained by
\begin{equation}
F_{V,A}^{2}=\frac{1}{g_{5}^{2}}\left(\frac{\Phi_{V,A}^{''}(0)}{N}\right)^{2},
\end{equation}
where
\begin{equation}
N= \int_{0}^{z_{IR}} dz  a \vert \Phi_{V,A}(z) \vert^2.\nonumber
\end{equation}
and the mass of the pseudoscalar meson can be obtained by solving
\begin{eqnarray}
&&\left( \partial_{z}^{2} + \partial_{z}\left(\ln a \right) \partial_z \right) \phi^a +g_{5}^{2} a^2 \left( M_{A}^{2} \right)_{\alpha\alpha} \left(\pi^{\alpha} - \phi^{\alpha} \right) =0, \nonumber\\
&&\partial_{z} \left( a^3 \left( M_{V}^{2}
+M_{A}^{2}\right)_{\alpha\alpha}
\partial_z \pi^{\alpha} \right) =
 q^2 a^3  \left( \left(M_{V}^{2} +M_{A}^{2} \right)_{\alpha\alpha} \left({1\over 2} \phi^{\alpha}-\pi^{\alpha} \right) +\left( M_{A}^{2} -M_{V}^{2} \right)_{\alpha\alpha} {1 \over 2} \phi^{\alpha} \right),\nonumber\\
\end{eqnarray}
with boundary
conditions:\\
$\partial_{z}\phi^{\alpha}(z=z_{IR})=\partial_{z}\pi^{\alpha}(z=z_{IR})=\phi^{\alpha}(z=0)=\pi^{\alpha}(z=0)=0$,
where $\phi^{\alpha}$ is defined as the longitudinal part of
$A_{\mu}^{\alpha}$,$~\partial_{\mu}\phi^{\alpha}=A_{\mu||}^{\alpha}$~.

The decay constants of the pseudoscalar
\begin{equation} f_{P^{\alpha}}= - {1 \over
g_{5}^2} {\partial_{z} A^{\alpha}(0,z) \over z}
\arrowvert_{z=\epsilon},
\end{equation}
with $A^{\alpha}(0,z)$ are given by the solution of
Eq.(\ref{eq.phiA}) satisfying $A^{\alpha \prime}(0,z_{IR})=0$ and $
A^{\alpha}(0,\epsilon)=1$.

In order to generate a mass gap, we need to introduce an IR
cutoff($z_{IR}$). The fifth dimension is taken as an interval from
$0$ to $z_{IR}$.

The model now has six free
parameters:$~g_{5}^{2}$,$~z_{IR}$,$~m$,$~c$,$~m_s$,$~c_s$.
$g_{5}^{2}$ can be obtained by comparing the vector-vector two point
function obtained from the OPE of QCD to that obtained using the
holographic recipe\cite{Erlich}, giving $g_5^2=\frac{N_c}{12\pi^2}$.
Thus there are five free parameters left, we use an iterative method
to fit the five free parameters.

We start by fitting the parameters without the back reaction. That
is, as a starting point we choose $a(z,m,c,m_s,c_s)=\frac{1}{z}$.
Using the following experimental
data:$~m_{\pi}=139.6MeV,f_{\pi}=92.4MeV,m_{\rho}=775.8MeV,m_{K1A}=1339MeV$,
and a semi-global fit for $m_{K^{*}}$, we then use an iterative
search method to fix the free parameters in order to minimise the
rms error on the remaining data. The final fit results are shown in
Table \ref{tab:parameter}.
\begin{table}
\tabcolsep 3mm \caption{Fit results for the free parameters in units
of MeV.}
\begin{center}\label{tab:parameter}
\begin{tabular}{|r|r|r|r|r|r|}
\hline $z_{IR}^{-1}$ & m &$ c^{\frac{1}{3}}$& $m_s$ &$c_s^{\frac{1}{3}}$\\
\hline 320.55&2.28&328.5&138.5&176\\
\hline
\end{tabular}
\end{center}
\end{table}
Having fixed the free parameters, we can calculate the remaining
mesons masses and decay constants. In Table \ref{tab:vector} and
Table \ref{tab:axial}, we show the mass and decay constants of
vector mesons and axial vector mesons respectively.

\begin{table}
\tabcolsep 3mm \caption{Axial vector meson results calculated with a
back reacted geometry and the free parameters given in Table
\ref{tab:parameter}. $*$ indicates that this value is used to fix
the free parameters, all other values are predictions. Numbers in
brackets give the percentage error.}
\begin{center}\label{tab:vector}
\begin{tabular}{|r|r|}
\hline observation&$value(MeV)(\% error)$\\
\hline $m_{\pi}$&139.6*\\
\hline $f_{\pi}$&92.4*\\
\hline $m_{a1}$&1364(10.9)\\
\hline $\sqrt{F_{a1}}$&440(1.6)\\
\hline $m_{K_{1A}}$&1339*\\
\hline $\sqrt{F_{K_{1A}}}$&435(4.1)\\
\hline $m_{A3}$&1344\\
\hline $\sqrt{F_{K_{A_{3}}}}$&412\\
\hline
\end{tabular}
\end{center}
\end{table}

\begin{table}
\tabcolsep 3mm \caption{Axial vector meson results calculated with a
back reacted geometry and the free parameters given in Table
\ref{tab:parameter}. $*$ indicates that this value is used to fix
the free parameters, all other values are predictions. Numbers in
brackets give the percentage error.}
\begin{center}\label{tab:axial}
\begin{tabular}{|r|r|}
\hline observation&$value(MeV)(\% error)$\\
\hline $m_{\rho}$&775.8*\\
\hline $\sqrt{F_{\rho}}$&348.8(1.1)*\\
\hline $m_{\rho^{'}}$&1781\\
\hline $\sqrt{F_{\rho^{'}}}$&658\\
\hline $m_{K^{*}}$&812(9)\\
\hline $\sqrt{F_{K^{*}}}$&328(11)\\
\hline $m_{V_{3}}$&$m_{\rho}$\\
\hline $\sqrt{F_{V_{3}}}$& $\sqrt{F_{\rho}}$\\
\hline
\end{tabular}
\end{center}
\end{table}
Having fitted the free parameters and calculated the remaining meson
properties we can also calculate the Ricci scalar for the back
reacted geometry:
 \begin{equation}
 R(z)=-12 c^{2}z^{6}-6 c_s^{2}z^{6}-8c z^{4} m-4c_s z^{4}
 m_s+\frac{5}{16}\big(8+2c^{2}z^{6}+c_s^{2}z^{6}+2c z^{4}m+c_s
 z^{4}m_s\big)^{2}.
 \end{equation}

\begin{figure}[ht]
\begin{center}
\includegraphics[width=8cm,clip=true,keepaspectratio=true]{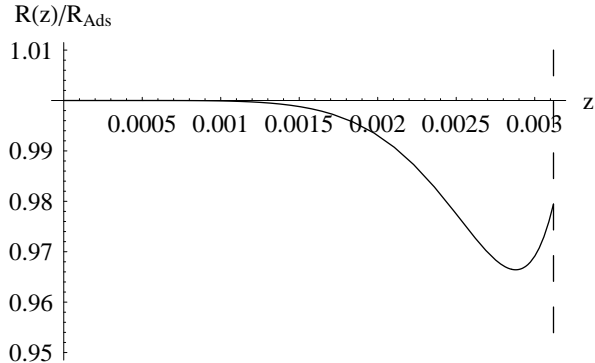}
\caption{The variation of the Ricci scalar as a function of radial
distance in our model with with parameters in Table
\ref{tab:parameter}.}
\end{center}\label{fig:ricci}
\end{figure}

\begin{figure}[ht]
\begin{center}
\includegraphics[width=12cm,clip=true,keepaspectratio=true]{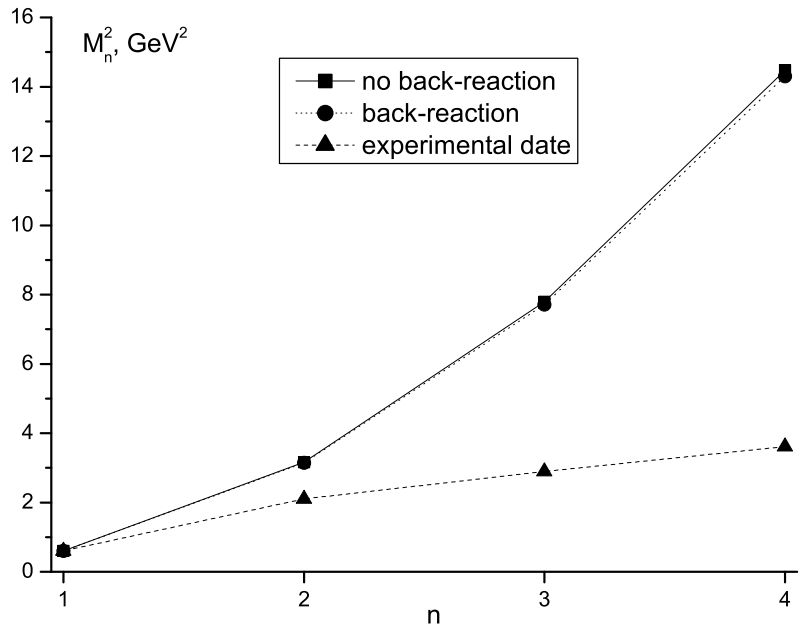}
\caption{Experimental values for $M_{\rho^{*}}^{2}(GeV^{2})$ against
theoretical values both with and without back-reaction. The results
without back-reaction are calculated with $z_{IR}^{-1}=322.6MeV$ and
those with back-reaction are calculated using the parameters given
in Table \ref{tab:parameter}. }
\end{center}\label{fig:resonance}
\end{figure}

We plot the curvature as a function of the radial distance in the
AdS space, see Figure 1. Figure 1 shows that the back reaction has
only a small impact on the scalar curvature in the interval
$(0,z_{IR})$ with a maximum of around $3\%$ departure from the pure
AdS result. For $z>z_{IR}$ the impact is larger but has no effect on
our results.

We also calculate the mass of $\rho$ resonances which is shown in
Figure 2. However, because of the small difference between
no-back-reaction case and back-reaction case, the two lines are
almost indistinguishable on this scale. As the stringy effects are
neglected in our present analysis, they are expected to become
important in the UV. Thus, the reliability of the current models
will diminish above the scale of chiral symmetry breaking (around
1200 MeV).

The main conclusion of this calculation is that even with the
addition of strange quark dynamics the geometry and hence the
spectra of masses and decay constants are not heavily affected. This
is a non-trivial statement about the impact of the strange quark on
chiral dynamics.

\acknowledgments We are grateful to Dr.Jonathan P. Shock and Dr.
Feng Wu for helpful discussions. This work was supported in part by
the National Science Foundation of China (NSFC) under the grant
10475105, 10491306 and the Project of Knowledge Innovation Program
(PKIP) of Chinese Academy of Sciences.

\end{document}